# Digital Twins solve the mystery of Raman spectra of parental and reduced graphene oxides

Elena F. Sheka

Institute of Physical Researches and Technology,
Peoples' Friendship University of Russia (RUDN University), 117198 Moscow, Russia

sheka@icp.ac.ru

**Abstract.** A still amazing identity of the D-G doublet Raman spectra of parental and reduced graphene oxides is considered from the digital twins' viewpoint. About thirty DTs, presenting different aspects of the GO structure and properties, were virtually synthesized using atomic spin-density algorithm, which allowed reliably displaying reasons for this extraordinary spectral feature. In both cases, it was established that the D-G doublets owe their origin to the $sp^3$-$sp^2$ C-C stretchings, respectively. This outwardly similar community of the doublets' origin of GO and rGO is thoroughly analyzed to reveal different grounds of the feature in the two cases. Multilayer packing of individual rGO molecules in stacks, in the first case, and spin-influenced prohibition of the 100% oxidative reaction, the termination of which is accompanied with a particular set of highly ordered by length $sp^3$- and $sp^2$ C-C bonds, protecting the carbon carcass from destruction caused by the stress induced $sp^2$-to-$sp^3$ transformation, in the second, are the main reasons. The DT concept has been realized on the basis of virtual vibrational spectrometer HF Spectrodyn.

## 1. introduction

Currenly, graphene oxide (GO) and reduced graphene oxide (rGO) top the list of high-tech modern graphene material science, which is explained by the relatively easy manufacturing, which tightly couples these materials, and their moderate cost. The list of publications is practically countless as can be seen from the recent reviews (see [1-9], but a few). Belonging to extensive family of solid carbons, GO and rGO live their lives quite separately. As occurred, GO does not exist in nature and is a synthetic product known since 1859 [10]. In contrast, rGO exists in nature for many millions of years in deposits of diverse $sp^2$ amorphous carbons, including various coals, shungites, anthraxolites, and accompanying carbon sheaths-shells of many other minerals [11]. However, the attribution of this richness to that of rGO has been realized only recently, when it was found that all the $sp^2$ amorphous carbons listed above are multilevel structures based on basic structural units (BSUs), which are graphene domains surrounded with necklaces of heteroatoms. This presentation of necklaced graphene molecules (HGMs) suits exactly the manufactured rGO as well, solids of which is just such a multilevel amorphous structure. Appealing to synthetic manufacturing [12-19], the modern graphenics classifies OG and rOG as an almost inseparable pair, although Nature provides huge arrays of natural products.

One of the motivations inclining to the acceptance of this internal kinship inseparability is the hitherto unsolved riddle of the amazing identity of the Raman spectra of GO and rGO, , whatever its form it may concern. If we take into account that the carbon backbones carcass of these two covalent carbons have completely different structure, which is a net of condensed non-planar cyclohexanoid units of GO in contrast to flat benzenoid ones of rGOs, then the observed identity becomes a unique spectral phenomenon, never before observed for covalent molecules and solids. At the same time, the high demand for both materials in modern high-tech production forces us to unravel this mystery, which gives rise to great uncertainty in understanding the internal structure of both materials, which evidently prevents their optimal use. This paper is aimed at solving the task attempting to answer the main question why Raman spectra of the two carbon oxides are so similar. It is precisely the lack of understanding of this exceptional oddity that forces material scientists to operate with a complex "identity certificate" of of these materials, figuratively presented in Figure 1 and based on a comparison of the properties of GO and rGO. As clearly seen, the structure and chemical composition of GO and rGO are completely different and the only characteristic that is the same for both substances is their Raman spectra.

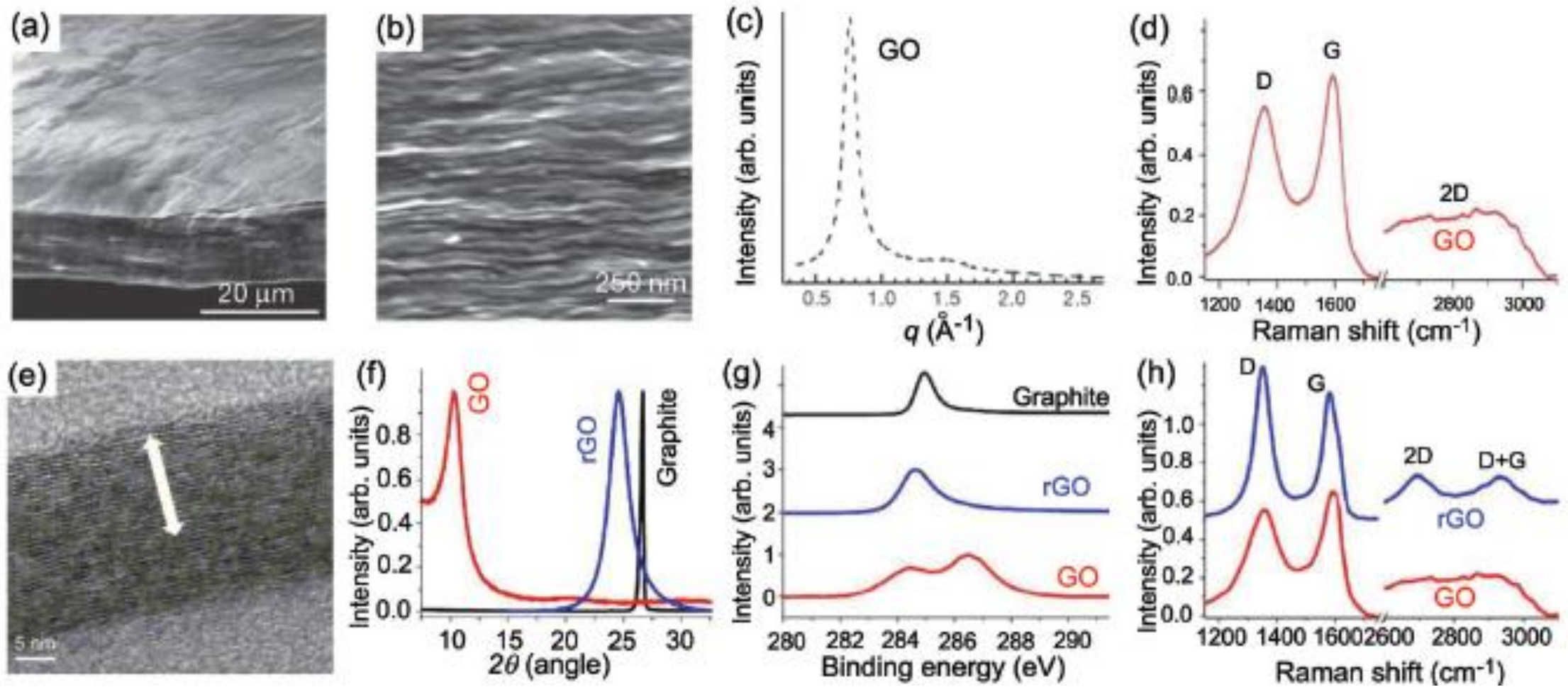

**Figure 1**. (a) Low- and (b) high-resolution SEM side-view images of a 10 mm-thick GO paper. (c) X-ray diffraction pattern of the GO paper sample. (d) Raman spectrum of a typical GO paper. (e) Cross-section TEM images of a stack of rGO platelets. (f) Powder XRD patterns of graphite, GO, and rGO. (g) XPS characterization of rGO platelets. (h) Raman spectra of rGO (blue) and the GO reference sample (red). Reproduced with permission from Ref. 6.

Spectroscopists all over the world are well aware that Raman spectra are widely used for structural and chemical analysis of substances, particularly covalen ones, precisely because of their dependence on both parameters. Naturally, the observed exotic identity could not fail to attract the attention of researchers. However, until recently this unique circumstance has not received a convincing explanation [20-22]. According to the author, two reasons underlie this situation. The first concerns the fact that material scientists are fixated on the Raman spectrum as the main indicator allowing to attribute the manufactured or used material to the group of graphene materials. Actually, numerous Raman spectra of various carbonaceous materials convincingly testify the presence of a characteristic D-G-band-doublet signature (see review [23] and references therein), which together with an independent evidence of planar graphene domains in the ground of the relevant carbon carcass allowed making undeniable verdict on the graphene nature of the bodies attributed as rGOs. Simultaneously, numerous efforts have been made to preserve the identification ability of this spectral mark, , which led to giving an exclusive role to two parameters of the spectra, namely the ratio of the intensity of the D and G bands, ${I_D}/{I_G}$, and the corresponding bands half-widths, $\Delta\omega$, to characterize size and defect structure of the relevant graphene domains. This relationship was established theoretically for graphene crystal [24,25] and then transferred to nanoscale rGOs of amorphous substances [26, 27]. However, as was shown lately [23], such a transfer turned out to be incompetent, which, nevertheless, has not stopped the efforts of the "theoretical description of the defectiveness" of the studied rGOs until now and has been accepted for GO by default.

The second reason for keeping the identity of the Raman spectra of GO and rGO uncovered is closely related to the first launch of the mandatory proof of the "graphene-domain origin" of the Raman spectrum of GO (see review [20] and references therein). This formulation of the problem was stimulated with the presence of 'zones of the original graphite that did not undergo oxidation' among the solid GO massive observed experimentally [28, 29]. In addition to the fact that the conclusions made in these articles raise serious doubts, it should be obvious that this circumstance, once taken place in the case of a low oxygen content in the GO mass, decreases in importance as the oxidation proceeds. However, the Raman spectra of GO show no traces, which might indicate their dependence on the oxygen content. Thus, a simplified idea appeared about weak interaction of graphene domains in GO with groups of atoms containing oxygen [30]. However, not only is the chemical action leading to the transformation of benzene into cyclohexane can hardly be considered weak, it is obvious, that an attempt to explain the 13 cm$^{-1}$ blue shift of the G bands in the GO spectrum compared to that in the rGO one, as the only difference between the bodies spectra, cannot be considered as a serious confirmation of the proposed explanation being applied to bands with a half-width of more than 100 cm$^{-1}$. Thus, the mystery of the identity of Raman spectra of GO and rGO remains unsolved. In turn, this problem is the cause of great confusion in the designation of the material used in a large number of articles devoted to 'graphene' application in the field of material science associated with various fields of chemical and biophysical technologies, as is, say the case of Covid vaccines [31].

A solution of the first part of the problem, which concerns general regularities governing IR and Raman spectra of rGO, has been recently suggested [32]. Accordingly, the main motives, laying the foundation of IR and Raman spectra of this class of solid carbons and providing strong and weak dependence of the spectra, respectively, on the body origin, were disclosed. The applied virtual vibrational spectrometry, based on the Digital Twins concept [32-34], occurred quite efficient and fruitful. As shown, the Raman spectra signatures contain information concerning spatial structure of individual graphene domains and their packing, while the molecules necklaces are responsible for IR spectra. Suggested sets of general frequency kits facilitate the detailed chemical analysis, providing a reliable basis for express analysis of any representative of the class. Our challenge now is to extend the obtained success to GO as well.

According to the concept, digital twins (DTs), which are molecular models, whose chemical composition and spatial structure reflect characteristic features of the substance under consideration, are objects of study in their own right. The implied possibility of a wide manipulation of the DTs, consciously aimed at strengthening one or another particularity of the structure or chemical composition of the real object, is the main feature of the approach, which makes it possible to establish the true state of affairs regarding the selected object, as a result of a broad comparative analysis. The goal of the virtual experiment is to reveal all the secrets of the dynamic behavior of matter and to obtain a convincing explanation of the structure of the Raman scattering and IR absorption spectra of GO. Presented below reveals achievements of the goal and obtaining a complete explanation of all the peculiarities of IR and Raman spectra of GO.

The paper is composed as follows. General grounds of the digital twins design alongside with the description of the first approach DTs and their vibrational spectra are given in Section 2. Section 3 concerns the design and virtual spectra of DTs of the second and third approaches, the latter particularly aided at monitoring $sp^2$-to-$sp^3$ transformation of the carbon carcass. Conclusive discussion of the digit twins concept and remarks concerning the exceptional aptitude of crbon atoms for this transformation are presented in Section 4.

## 2. Digital twins of graphene oxide

### 2.1. General grounds of the digital twins design

Both GO and rGO are polyderivatives of bare honeycomb graphene domains, but of different class. When rGO presents a large family of necklaced graphene molecules, GO is a product of the derivatization that includes not only edge carbon atoms of the domain, but basal-plane ones as well. Once derivatized, both edge and basal-plane $sp^2$ carbon atoms of the pristine domains undergo a transition to $sp^3$ hybridization. However, in the case of rGO, this $sp^2$-$sp^3$ transformation, concerning edge atoms only, may not occur, retaining $sp^2$ atom character provided with the addition of individual hydrogen or fluorine atoms as well as hydroxyl or carboxyl units, thus giving $sp^2$ character to the relevant rGO as a whole. This usually happens in large massives of both natural and chemically produced $sp^2$ amorphous carbons, BSUs of which are rGO fiattened molecules based on $sp^2$hydridized carbon carcass [11]. In contrust, the derivatization of basal-plane atoms involves in the mandatory sp2-sp3 transformation of both edge and basal-plane atoms since it starts when a full transformation of the edge atoms is completed [35, 36]. Thus, such well known chemical derivatives of bare graphene as graphane [35] and/or graphene oxide [36] are $sp^3$ configured carbon materials. Returning to the formulation of the problem, now in the light of the information obtained, it can be changed as follows: we are faced with the task of obtaining an explanation for the identity of the Raman spectra of $sp^2$ rGO and $sp^3$ GO.

Basal plane oxidation results in a severe transformation of the plreviously flat benzenoid structure thus transfering the latter to a riffled cyclohexanoid one. From the vibrational dynamic viewpoint, benzenoid and cycloxexanoid vibrational signatures are absolutely different [37, 38]. Additionally, the cyclonoid structures are subjected to a large conformity, which greatly complicates both the prediction of the post-derivatization structure and vibrational analysis of the latter (see [35] and references therein). For the first time, on the computational level this problem was discussed when considering the hydrogenation of a (5,5) graphene domain ((5,5)NGr below) [35] and then its oxidation [36]. One of the ways to clarify what is going with the pristine domain under edge-and-basal derivatization is to trace subsequent steps of the reaction one by one. Computationally, it was carried out by applying the spin molecular theory of $sp^2$ nanocarbons [39], perfectly demonstrated its efficacy by virtual design of the graphene derivatives mentioned above as well as polyderivatives of fullerene $C_{60}$ under stepwise fluorination and hydrogenation and so forth [40-42]. The virtual synthesis is followed a spin-density algorithm, in frame of which, the value of the atomic chemical susceptibility (ACS) is a quantitative indicator of the atom chemical activity allowing to match targets for any next step of the considered reaction.

The algorithm is based on the ACS dependence on the length of the relevant covalent C-C bond [43]. Concerning $sp^2$C-C bonds, ACS is nil until the bond length overcomes the critical value $R_{crit}$ = 1.395 Å. As occurred, usually UHF calculated bond lengths of the graphene domain fill the interval of 1.322-1.462 Å. The relative number of bonds, whose length exceeds $R_{crit}$, constitutes 62% of the bond massive The resulting radicalization leads to a considerable amount of the effectively unpaired electrons, total numbers of which $N_D$ constitutes 31 e, in the case of (5,5) NGr. From the chemical viewpoint, $N_D$ value describes the domain (molecular) chemical susceptibility (MCS), while $N_{DA}$ presents ACS related to atom A. These two quantities are main parameters of the spin-density algorithm of the graphene domain derivatization and simultaneously exhibit local spins and their distribution over the domain atoms [44].

Figure 2a exhibits a typical ACS image map presented by the $N_{DA}$ distribution over 66 atoms of the (5, 5)NGr. The molecule edges are not terminated, and the ACS map has a characteristic view with a distinct framing by edge atoms since the main part of the unpaired electrons is concentrated in this area. This map presents a typical chemical portrait of any graphene domain and exhibits the exceptional role of the circumference area, which looks like a typical 'dangling bonds' icon. At the same time, the ACS image map intensity in the basal plane is of ~0.3 *e* in average so that the basal plane should not be discounted when it comes to chemical modification of the molecule thus disproving the common idea of its chemical inertness. Moreover, basing on the $N_{DA}$ value and choosing the largest of them as a quantitative pointer of the target atom for the coming chemical attack, one can perform a reliable virtual synthesis of the domain polyderivatives [39].

The absolute $N_{DA}$ values of the (5, 5) NGr, shown by light gray plotting in Figure 2e, clearly exhibit 22 edge atoms involving 2x5 *zg* and 2x6 *ach* ones with the highest $N_{DA}$ thus marking the perimeter as the most active chemical space of the molecule, successively terminated one by one firstly. The first step of the derivatization occurs on atom 14 (see star-marked light gray plotting in Figure 2e) according to the largest $N_{DA}$ in the output file. The next step of the reaction involves the atom from the edge set as well, and this is continuing until either all the edge atoms are saturated or some of the basal ones come into play. In the case of hydrogenation, all 44 steps are accompanied with the high-rank $N_{DA}$ list where edge atoms take the first place thus being terminated by hydrogen pairs [35]. Thus obtained hydrogen-necklaced (5, 5)NGr molecule is shown in Figure 2c alongside with the corresponding ACS image map in Figure 2d which reveals the transformation of brightly shining edge atoms in Figure 2b into dark spots. The addition of two hydrogen atoms to each of the edge ones saturates the valence of the latter completely, which results in zeroing $N_{DA}$ values as is clearly seen in Figure 2e. Thus, the chemical activity is shifted to the neighboring basal atoms and retains higher near *zg* edges. Dotted curve in the figure exhibits free valence distribution over the molecule atoms in basal plane (see details in [39]). The hydrogenation of atoms in the basal plane of the necklaced (5, 5) NGr starts on atom 13 (see star-marked black plotting in Figure 2e). Evidently, the order in which carbon atoms come into per-step play depends on the chemical reagent, the fixing conditions for the graphene domain, as well as on *up*-and/or-*down* accessibility of the basal plane atoms. These issues are discussed in details with respect to the hydrogenation and oxidation of the (5, 5) NGr domain [35, 36, 45]. Three graphene hydrides (GHs) and three GOs were synthesized this way. Important to note, that the ACS-algorithmic per-step synthesis solves the problem of the $sp^2$-$sp^3$ transformation of the carbon bond structure of rGO into that one of GO in the most natural way, thus allowing to trace where and when the quantity turns into quality and where along this path the identity of the Raman spectra of rGO and GO appears. This highly efficient method of virtual synthesis of graphene domain derivatives was laid into the foundation of the DT design in the current study.

## 2.2. Digital twins of the first approach

Three GOs previously synthesized by our team [36] form the basis of the DTs of the first approach. Their equilibrium structure is shown in Figure 3. The basic (5,5) NGr is of 1,2x1,1 $nm^2$ in size and suits well not only empirical necklaced-graphene compositions of rGO related to $sp^2$ amorphous carbons [32], but to $sp^3$ structural units of GOs of ~2 nm$^2$ in size [46]. Nothing to say that the identity of graphene domain in the basis of these DTs related to both GO and rGO, is exeptional favoring bonus for the current study.

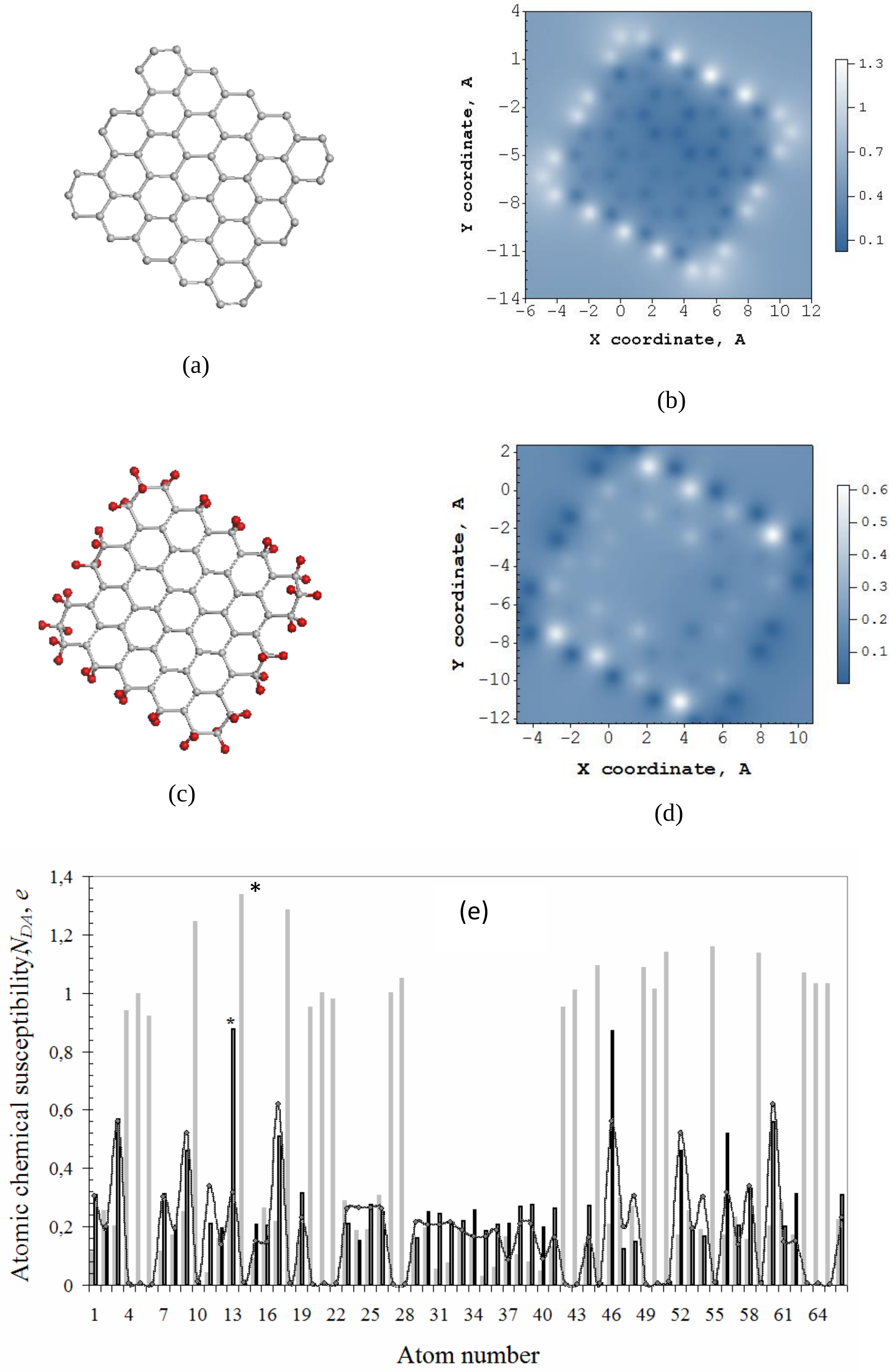

**Figure 2**. Equilibrium structures of pristine domain (5, 5) NGr $C_{66}$ (a), free-standing (c) (5, 5) NGr, terminated by two hydrogen atoms per each edge one $C_{66}H_{44}$, and ACS $N_{DA}$ image maps over atoms in real space (b, d) as well as according to the atom number in the output files (e). Light gray histogram plots ACS data for $C_{66}$, while black one and curve with dots are related to $C_{66}H_{44}$. The curve maps the free valence distribution over atoms. Scale bars match $N_{DA}$ values. UHF AM1 calculations.

Pronounced isomorphism of cyclohexanoid compositions determines the dependence of the formed structure on such external factors as the fixation or free standing of the graphene domain edges, the accessibility of the domain basal plane to heteroatoms participating in the reaction from one or both sides (*up* and *down* format). The determining role of these factors in the formation of the final product can be traced on the example of the virtual synthesis of graphene hydrides [35]. In what follows, a particular marking subscriptions *fx*, *fr*, *1*, and *2* will be used to distinguish external conditions of the DTs formation. The set of GO isomorphs GO1, GO2, and GO3 in Figure 3 forms a series *fr2*, *fr1*, and *fx2* indicating that the graphene domain edges of the first two were free standing while those ones of GO3 were fixed. Simultaneously, the basal plane in the case of GO1 and GO3 was accessed *up* and *down*, while in the DO1 case the accession was one-side only. Concerning the current case, the virtual synthesis is complicated with the difference in oxygen-containing units (OCUs) involved [36]. GO1 and GO2 are the result of oxidation in the simultaneous presence of three such groups, namely, atomic oxygens, hydroxyls and carboxyls. All the groups were considered at each synthetic step and the prevalence to be selected for the synthesis continuation was given by the biggest energy of the derivative formation. The third GO3 member presents a product, virtually synthesized in a flow of hydroxyls only. The presented DTs are the result of about 400 computational jobs [36].

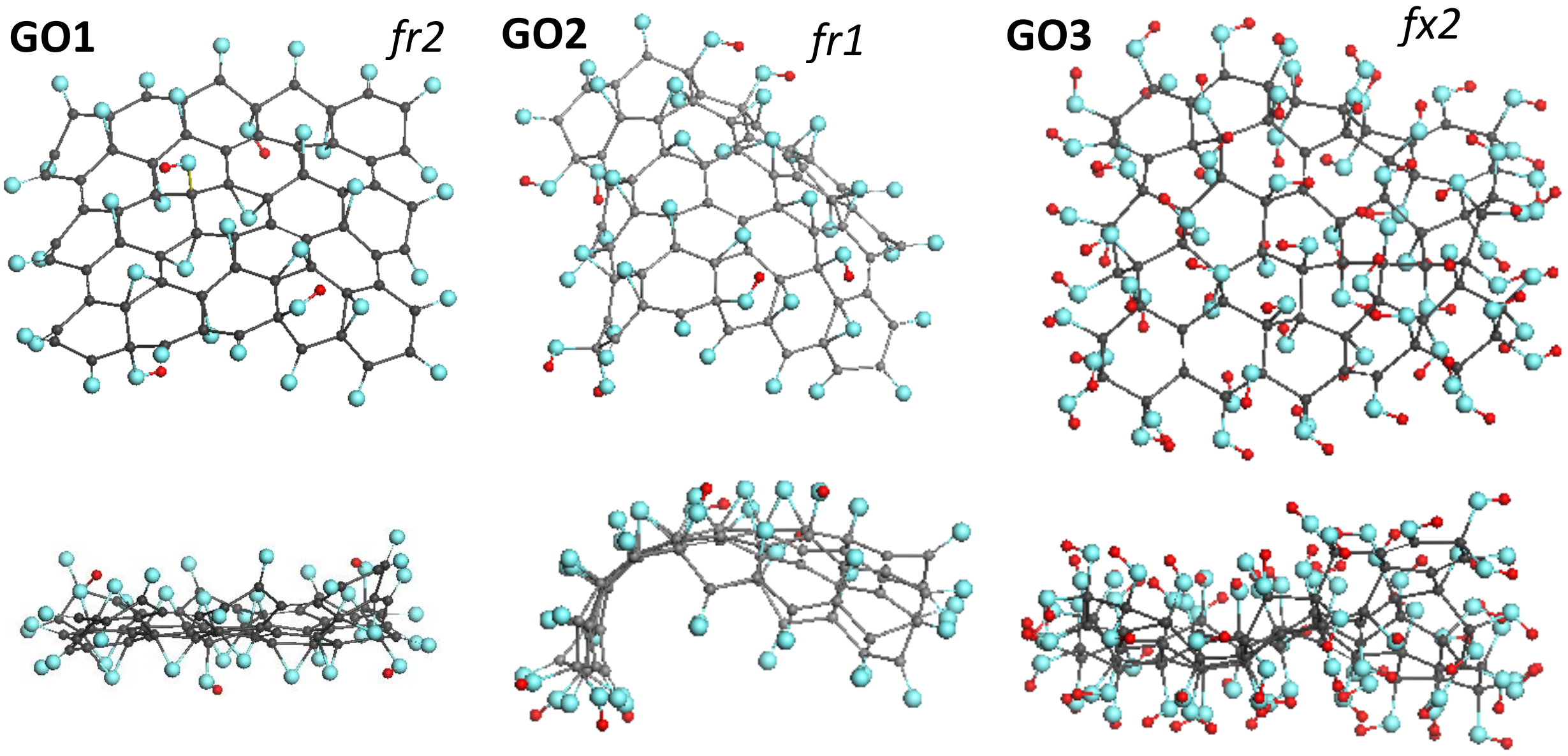

**Figure 3**. Face and side view of equilibrium structures of virtually synthesized graphene oxides GO1 and GO2, both of $C_{66}O_{40}H_4$ content, and GO3 of $C_{66}(OH)_{74}$. Gray, blue, and red balls mark carbon, oxygen, and hydrogen atoms, respectively. UHF AM1 calculations.

In the course of the virtual reactions, it was found that carboxyls are characterized by the least formation energy at each step due to which their presence in the basal plane area is highly likely. As for the graphene domain circumference, some single attachments might be possible under particular conditions of strong perturbation of the structure. This conclusion is supported with GO experimental structural data. As shown, solid GO consists of layered stacks of a few nm in thickness and of 2nm$^2$ in lateral dimension [28, 46]. The interlayer distance drastically depends on the water contamination, caused by the solid high hydrophilicity, and the minimal value constitutes 0.784 nm [47, 48]. If take into account that the thickness of a cyclohexanoid is bigger than that of benzenoid on 0.124 nm, carbon atoms take on 0.459 nm. Remaining 0.325 nm allow the location of either oxygen atoms (0.304 nm [49]) of epoxy groups or bent hydroxyls between the layers. The interlayer distance is rather small as well to comfortably house carboxyls in the circumference of the GO basic units. The groups are quite cumbersome and their comfortable disposition around graphene domain requires a lot of space (see DT **IV** structure in [32]). In accordance with this, we believe that the widespread statement such as "Presence of hydroxyl and epoxy groups on the basal plane of sheets and carboxyl groups at their edges opens the way for chemical modification of as-prepared GO" [9], based on first concepts of the chemical composition of graphene oxide [50, 51], discords with the reality. Accordingly, carboxyls will not be considered in this work when considering DTs design.

### 2.3. Virtual vibrational spectra of the digital twins of the first approach

The virtual device HF Specrodyn provides the calculation of harmonic one-quantum spectra of IR absorption and Raman scattering of preliminary structurally optimized DTs [52]. The calculations are based on the standard rigid-rotor harmonic-oscillator model in the framework of semiempirical Hartree-Fock quantum chemical approach. A detailed description of the code is given elsewhere [53]. All the calculations, discussed in the current paper, were performed using either UHF or RHF versions of the code AM1 approximation depending on the radical status of the studied DTs (see a detailed discussion of the problem in [32, 52]). Through over the paper, the virtual spectra are presented by stick-bars convoluted with Gaussian bandwidth of 10 $cm^{-1}$. Intensities are reported in arbitrary units, normalized per maximum values within each spectrum. Since the number of vibrational modes, composing the spectra under consideration, is too large, the excessive fine structure, statistically suppressed in practice, is covered by trend lines averaged over 50 next steps of linear filtration of 0.003÷0.010 $cm^{-1}$, which is indicated in captions.

Virtual vibrational spectra of the DTs, discussed above and presented in Figure 3, are shown in Figure 4. As seen in the figure, both IR and Raman spectra of GO1 and GO2 are well similar and differ from those ones related to GO3. The subsequent description of these virtual spectra is based on the general frequency kits (GFKs) presented in Table 1. The table includes GFKs previously adapted for $sp^2$-configured rGOs [32], while supplemented with the data proposed for $sp^3$ GO on the basis of experimental spectra in the current study. The analysis was carried out on a virtual frequency scale unless otherwise indicated. IR spectra of GO1 and GO2 are presented at 2100 $cm^{-1}$ with the $\nu$ $sp^3$**C =O** mode, while those of $\nu$ $sp^3$**C**-**O**H, $\nu$ $sp^3$**C-O-C**, and $\delta$ *ip* $sp^3$**C**-**O**H comparatively equally contribute to the region of 1200-1700 $cm^{-1}$. In the GO3 IR spectrum, the main role is expectedly assigned to vibrational modes involving hydroxyls, which are presented with $\delta$ *op* $sp^3$**C**-**O**H mode in the region of 400-1000 $cm^{-1}$ as well as with a large set om modes, including $\delta$ *ip* $sp^3$**C**-**O**H, $\nu$ $sp^3$**C**-**OH**, $\nu$ $sp^3$**C**-**O-C**, $\nu$ $sp^3$**C**-**C**, and $\nu$ $sp^2$**C**-**C** ones in the region of 1200-1800 $cm^{-1}$. As seen from the table, in contrast to rGO, the $\nu$ $sp^2$**C**-**C** mode pool of which is distinctly separated by frequency, thus promoting a tight connection of necklace heteroatoms and basal-plane carbon ones in the formation of IR and Raman spectra, respectively [32], the $\nu$ $sp^3$**C-C** modes of GO overlap closely with others, thus excluding a similar distinguishing. The feature evidently greatly complicates a detailed vibration-mode analysis of both IR and Raman spectra of GO and exacerbates the surprise caused by the standard appearance of the Raman spectrum of the GO.

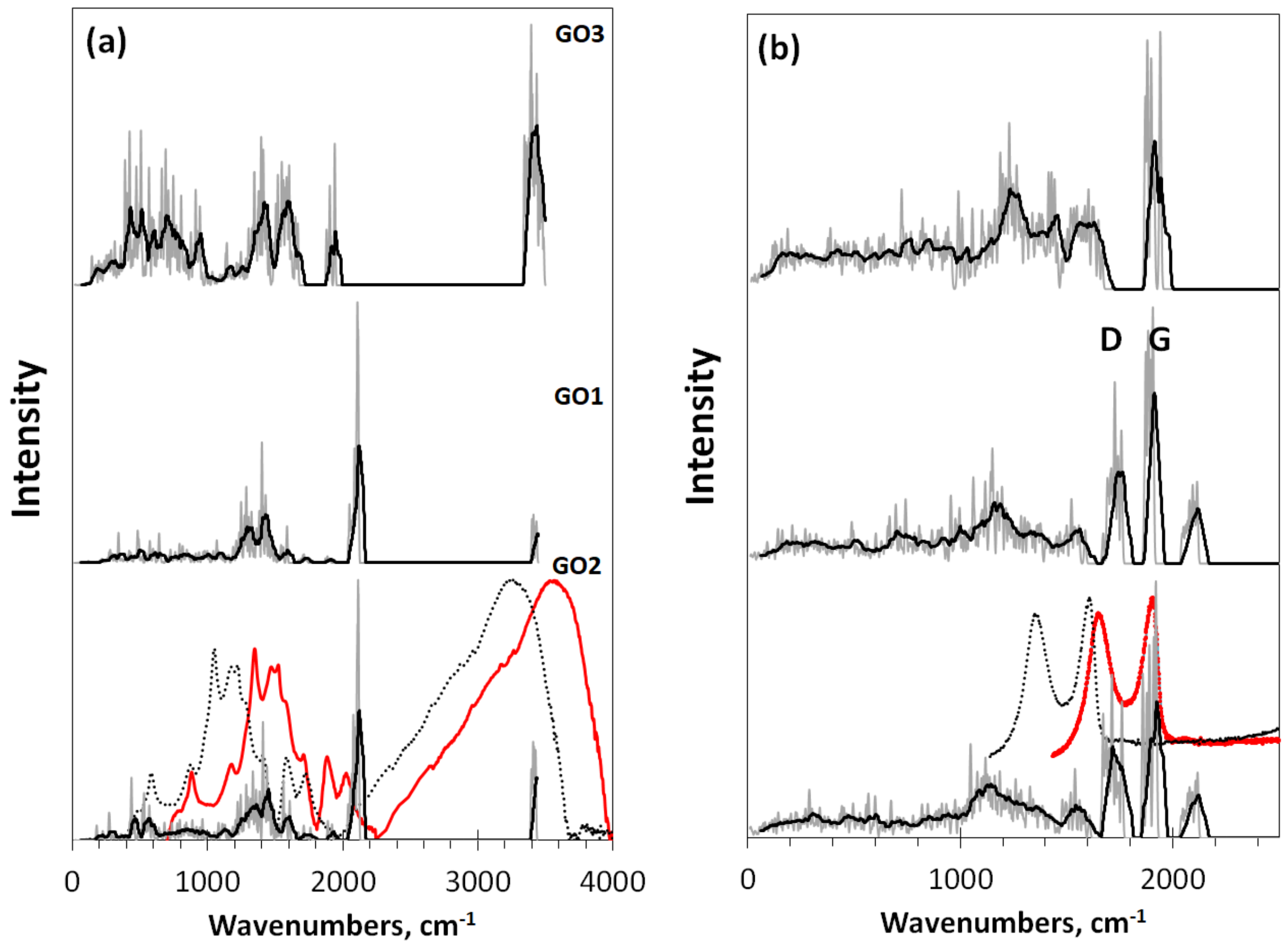

**Figure 4**. Virtual one-phonon IR absorption (a) and Raman scattering (b) spectra of digital twins GO1, GO2, and GO3. Spectra plottings are accompanied with trend lines, corresponding to 50-point linear filtration. RHF AM1 calculations. Dotted and red plottings present original and blue shifted on 300 $cm^{-1}$ experimental spectra of GO produced by AkKo Lab [60], respectively.

Looking at the Raman spectra of DTs GO1 and GO2 in Figure 4b, one immediately notices a triplet of intense bands at ~1760 $cm^{-1}$ (**I**), ~1900 $cm^{-1}$(**II**), and ~2100 $cm^{-1}$ (**III**). According to the GFKs listed in Table 2, the bands should be attributed to the $\nu$ *sp*$^3$**C**-**C**, $\nu$ *sp*$^2$**C**-**C**, and $\nu$ *sp*$^3$**C**=O modes, respectively. The latter two attributions are non-alternative while the first one can be mixed. The $\nu$ *sp*$^3$**C =O** origin of band **III** makes it possible to estimate the shift of virtual frequencies in this region with respect to the experimental ones. According to the data listed in Table 1, the blue shift is of 200-300 $cm^{-1}$. Determining that the largest value makes it possible to obtain a better agreement with the experimental data for the $\nu$ *sp*$^3$C-**O-H** band at 3400 $cm^{-1}$, we take this value as the desired one and apply it to the experimental spectra of a real sample produced by AkLab [60] (see a detailed description of the product and its spectra in [61]). As can be seen in Figure 4, shifting leads to an obvious agreement between the calculated and experimental spectra concerning both IR absorption and Raman scattering. In the latter case, the characteristic empirical doublet of D-G bands conveniently covers virtual bands **I** and **II**, making it possible to consider **I-II** bands the DTs of D-G ones. Naturally, we are not talking about exact replicas, but about agreement in the main elements of the structure, which allows their interpretation. Taking into account the amorphous nature of GO, one should expect the appearance of only the most significant structural elements in the experimental spectra, located against a considerable structureless background. Indeed, usually the experimental spectra have just such a character [21], as a result of which the doublet of D and G bands becomes the characteristic image of the GO Raman spectrum.

A general consistence of experimental and virtual IR spectra in Figure 4a alongside with a large set of GFKs listed in Table 1 allows making a confident detailed analysis of the IR spectrum structure of any real GO sample [61]. In this work, we will be interested only in the Raman spectra and in clarifying the question of whether bands **I** and **II** are really digital twins of the D and G ones. To get convincing answer, it is necessary to establish which namely $sp^3$ vibration modes lay the foundation of band **I** and to find the source of the $\nu sp^2$**C**-**C** origin of band **II**. Additionally, we should understand if the observed **I-II** doublet of the GO virtual Raman spectrum is stable enough to be a characteristic spectral signature of the body. As occurred, a design of a particular set of DTs made it possible to achieve the desired goal.

## 3. Monochrome digital twins of GO of the second approach

### 3.1. The problem of the GO polychromicity

Model oxides GO1 and GO2, shown in Figure 3, were synthesized ten years ago in an attempt to realize in their structure all the information about the chemical composition of real GO available by that time [36]. The domain (5, 5) NGr $C_{66}$ was taken as the basis for the synthesis, the first stage of which, involving the domain edge atoms, was accompanied by a successive consideration of the addition of single oxygen atoms, hydroxyls, and carboxyls at each step. After evaluating the binding energy (BE) of each attachment separately, the choice of the obtained graphene domain derivatives was made in favor of the configuration with the highest BE. For this derivative, its largest ACS value points the number of a target atom of the next oxidation step. This process continued with 22 steps, as a result of which the necklaced graphene molecule $C_{66}O_{22}$ was formed, and the oxidation process moved to the basal plane. Each step of this process was considered as a choice between the involvement of a $sp^2$C=C bond in the formation of either a $C_2O$ epoxy group or opening the bond with the addition of two hydroxyl groups thus forming $C_2(OH)_2$ composition. In addition, the landing of heteroatoms *up* or *down* in the case of GO1 and only *up* in the case of GO2 was controlled. The oxidation process was stopped at the 18$^{th}$ basal-plane concerning basal plane atoms due to zeroing the corresponding $N_D$ and $N_{DA}$ values. In both cases, the basal plane, consisting of 44 carbon atoms, is covered with 14 $C_2O$ epoxy and 4 OH groups.

Concerning chemical content and linear dimensions, GO1 and GO2 are well consistent with experimental data, which should be attributed to the basic structural units of real GO samples. Unexpectedly, their virtual Raman spectra presented in Figure 4b confirmed the validity of these models to an even greater extent implying that particular peculiarities of vibrational spectra of GO1 and GO2 do provide the characteristic **I-II** doublet Raman spectrum of real GO. The fact is highly exciting but reasons of such an exclusive behavior stays in the shadow. At the same time, the rigid restriction of the experimental Raman spectrum to just this doublet of bands and its identity to the rGO spectrum evidently indicates the presence of undoubted features in the vibrational spectrum of GO as well as of particular principles that determine the features manifestation in the Raman spectrum. However, the described oxygen status of GO1 and GO2 is quite complicated, too polychromic, and does not allow making any conclusions about the causes of these features beyond speculations. Evidently, the monochromization of the oxide composition may be a logical step towards overcoming this polychrome difficulty and solving the spectral riddle. The first attempt was made 10 years ago [36], as a result of which complex oxides GO1 and GO2 were replaced by oxide GO3. Its virtual synthesis, based on the (5, 5) NGr as well, was considered with the participation of hydroxyls only. At the first stage of the reaction, a $C_{66}(OH)_{36}$ necklaced graphene molecule was synthesized. After moving the reaction to the basal plane, another 38 OH groups were added *up* and *down*. As can be seen from Figure 4, the monochromization drastically affects both IR and Raman spectra. As for the latter, only the band **II** is observed in the Raman spectrum of the molecule, the position of which coincides with that one in the GO1 and GO2s spectra, while the band **I** disappeared. Understanding that the presence of the **I-II** doublet in the Raman spectra of GO1 and GO2 is provided with their oxygen polychrome content, a particular 'decomposition' of the latter is needed to isolate that component of this polychromatism, which determines the existence of bands **I** and **II**. It turned out to be possible to implement such a decomposition using a set of DTs specially configured.

3.2. Digital twins of the second approach

DTs related to GO consist of two structural parts concerning the oxygen content. The first involves the necklace atoms, linked to the edge atoms of the parental graphene domain and provided $sp^3$ hybridization of the latter in contrast to the $sp^2$ format in the rGO case [32]. The second part concerns configurations on the domain basal plane. Evidently, the influence of these two parts on IR and Raman spectra of GOs is different. DTs GO4, GO5, and GO6, shown in Figure 5, were designed to clarify this point. Based on (5, 5) NGr, GO4 and GO5 differ in the necklaces (oxygen atoms providing the carbonyl one in the first case and hydrogen pairs forming the methylene necklace in the second), once covered *up* and *down* with 22 epoxy groups in both cases. Against, the necklace composition of DTs GO5 and GO6 is the same, while epoxy covering of GO5 is replaced by hydroxyl one in GO6. The virtual synthesis, based on the ACS spin-density algorithm, concerned the necklaced parts of the oxides, while *up* and *down* filling of the domain basal plane was performed by hands.

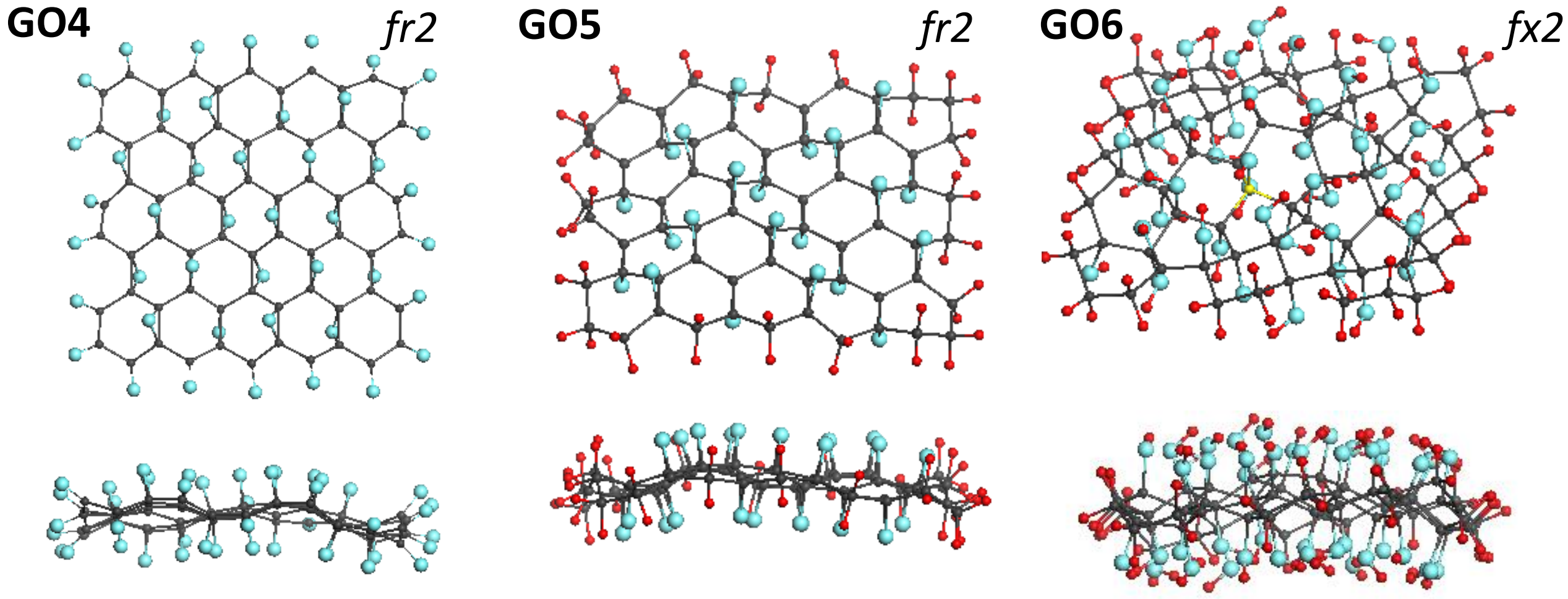

**Figure 5**. Face and side view of equilibrium structures of virtually synthesized monochrome graphene oxides GO4 ($C_{66}O_{44}$), GO5 ($C_{66}H_{44}O_{22}$), and GO6 ($C_{66}(OH)_{88}$). Gray, blue, and red balls mark carbon, oxygen, and hydrogen atoms, respectively. RHF and UHF AM1 calculations.

IR and Raman spectra of these DTs, shown in Figure 6, respond to the produced monochromatization in different ways. Concerning IR absorption, the replacement of carbonyl necklace of GO4 with methylene one of GO5 and GO6 results in a drastic changing of the spectra when going from GO4 to GO5. The next transformation of GO5 into GO6 evidently conserved the methylene character of the necklace, although noticeably influenced by events occurring in the basal plane of the carbon domain. Thus, as seen in Figure 6a, IR spectra exhibit vibrations of necklacing heteroatoms mainly, although influenced by the basal-plane events. In contrast, as seen in Figure 6b, Raman spectra of GO4 and GO5 are well similar with respect to the band **I** at 1760 $cm^{-1}$ already familiar to us while the band is absent in the GO6 spectrum. The situation is similar to that previously discussed for the GO1- GO3 spectra, which allows suggesting that the band appearance is tightly connected with the epoxy coverage of the domain basal plane, while a total hydroxylation of the plane does not suit the band origin. With regard to the spectra structure around band **I** and in accordance with the data of Table 1, the spectral area at 2200 $cm^{-1}$ is evidently attributed to $\nu\, sp^3$**C**=O vibration modes while those of 1000-1700 $cm^{-1}$ cover a large mixture of different modes among them $\nu\, sp^3$**C**-C stretchings should be considered first. As followed from the virtual vibrational spectrum of nanographane, based on (5,5)NGr [62] (see Figure 9 below), which is in a good consistence with virtual phonon spectrum of crystalline graphane [63], these modes are strictly confined in the frequency region of 1000-1500 $cm^{-1}$. As seen in Figure 6b, this confinement really takes place in the spectra of GO4 and GO5, while In the case of GO6, we see a clearly visible extension of the region up to 1700 $cm^{-1}$. Important to note, that therewith the extended high-frequency edge of the stretched spectrum of GO6 overlaps with the low-frequency wings of bands **I** in spectra of GO4 and GO5 thus, possibly hinting in this way at the common origin of the sections of the spectra under consideration.

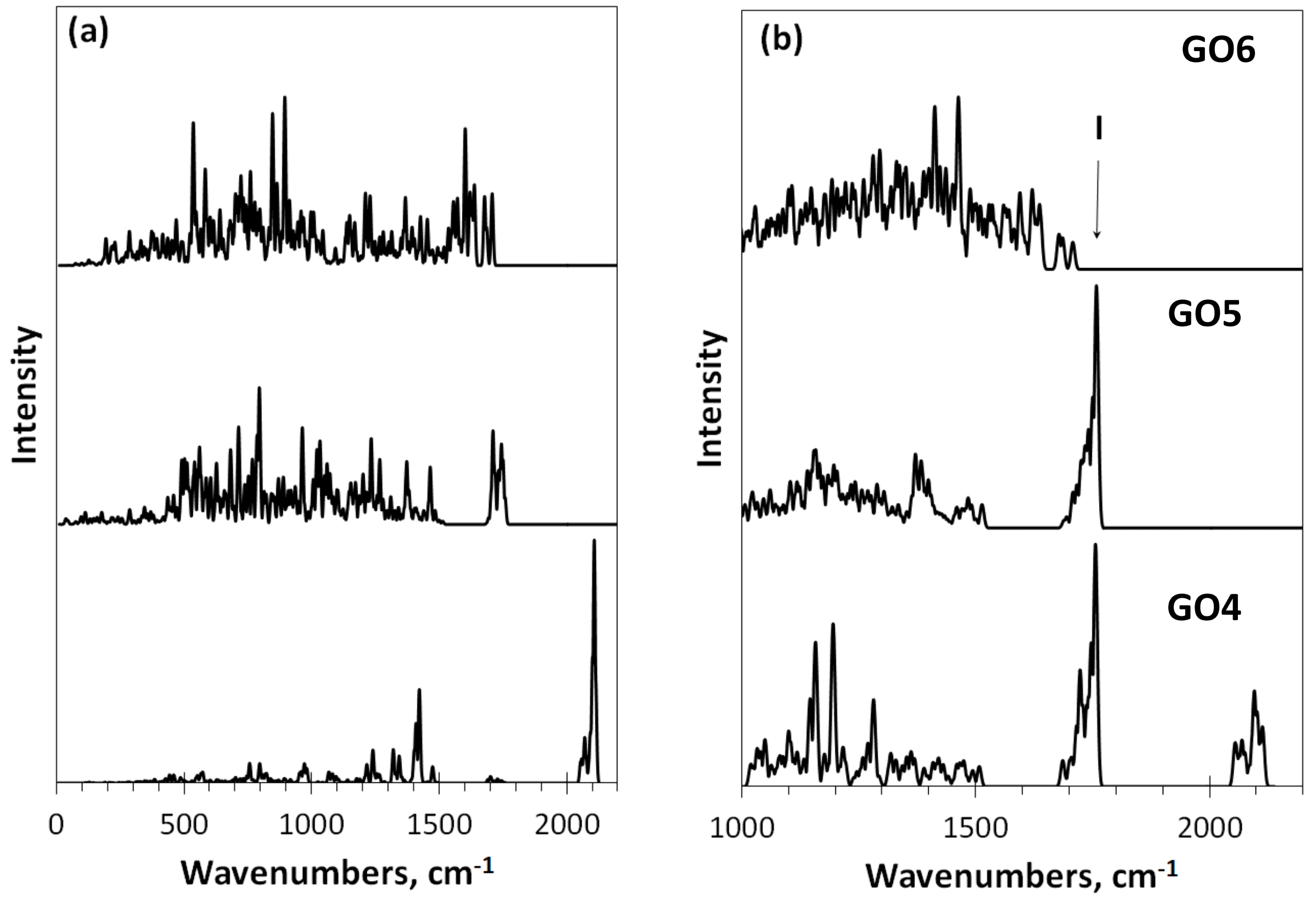

**Figure 6**. Virtual one-phonon IR absorption (a) and Raman scattering (b) spectra of digital twins GO4, GO5, and GO6. RHF AM1 calculations.

Since the carbon carcass of the current oxides is formed with the $sp^3$C-C valence bonds only, it should be recognized that these spectral intervals are due to the appearance of specific valence vibrational modes in the vibrational spectrum of oxides due to the presence of oxygen atoms. A peculiar behavior of vibrational bands in organic molecules associated with oxygen atoms was noticed as early as 40 years ago [64, 65] and since then has been constantly discussed in the literature (see [54, 55] and references therein). These bands easily changed their position and intensity depending on the place of these atoms in the molecule, on their number, on the presence of certain heteroatoms, and so forth. However, until now it does not go beyond the statement of the fact as well as attempts to link the latter with the presence of special valence bonds. In the case of the discussed oxides, which are oxygen-polytarget objects, nature itself provides us with a unique way to relate the observed spectral effect to the structure of valence bonds monitoring directly the $sp^2$-to-$sp^3$ transformation of the valence bonds of the pristine graphene domain. The possibility of such monitoring was demonstrated earlier on the examples of the virtual synthesis of $C_{60}$ fullerene derivatives [40-42] and a bare graphene domain (5,5) NGr [35, 36, 45]. Below we will apply this approach to GO4.

### 3.3. Monitoring o the $sp^2$-to-$sp^3$ transformation with DTs of the third approach

We consider the formation of GO as a gradual transformation of the $sp^2$ benzenoid carbon structure of graphene into $sp^3$ cyclohexanoid configuration of oxides in the presence of particular oxygen reagents. Previous investigations, discussed in Section 3.1, convincingly evidence that the ACS spin-density algorithmic approach, which ws in the grounds of the synthesized TDs, allows monitoring the oxidation step by step. The monitoring results can be fully presented with a set of data that include the relevant DT equilibrium structure, the distribution of its C-C covalent bonds over lengths, as well as its virtual Raman spectrum. We will use this very set below to isolate characteristic evidences of particular consequences of the above transformation. Skipping the first stage of the oxidation, which concerns the edge atoms of the basic domain (5,5) NGr and taking thus formed necklaced graphene molecule $C_{66}O_{22}$ (rGO, or GO4_00 below) the reference point of the basal-plane-carbon oxidation, Figure 7 exhibits results concerning the first four steps of the basal-plane oxidation. The choice of each subsequent-step target bonds is governed by the ACS spin-density algorithm in the format of the *up* and *down* basal plane access, applied to only one atom of the pair in contrast to previous synthesis [36] when both atoms were algorithmically selected.

As seen in the figure, the location of the first epoxy group on the basal plane (GO4_01) results in a significant reconstruction of at least 20 bonds of the reference $sp^2$C-C bonds pool, which is caused by the expected elongation of bonds 41, 43, 44,and 48. It should be noted that the bond number retains fixed throughout the subsequent oxidation cycle. Confirming that Raman scattering produces a spectral signature of the C-C bond configuration [32], the spectrum in comparison with the reference GO4_00 one [32], reveals a sharp response to the change in the bond structure. This feature accompanies the addition of each next epoxy group. Thus, the second epoxy group addition reveals the bond transformation of GO4_02 more markedly and this effect increases at the subsequent steps, and so forth. Since the observed $sp^2$-to-$sp^3$ transformation of the honeycomb structure is accompanied with unavoidable mechanical stress, the compensation of the latter causes not only expected elongation of the reference $sp^2$C-C bonds, but also its considerable shortening to protect the whole carbon body from destruction.

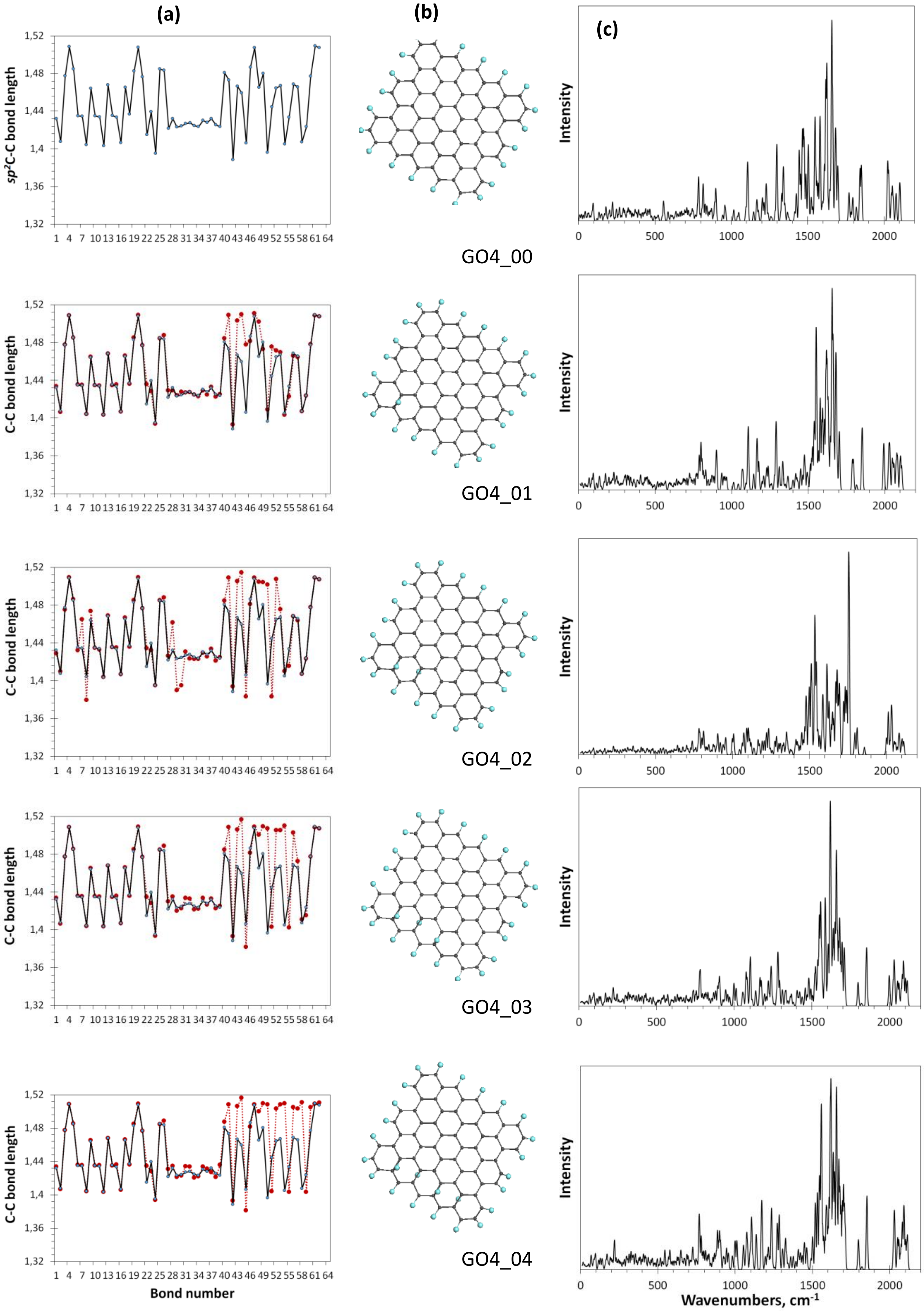

**Figure 7**. Monitoring of the $sp^2$-to-$sp^3$ C-C bond transformation at first steps of the oxidation of basal-plane carbon atoms of the reference GO4_00 TD. a. TD C-C bond length (reference and current presented with black and dotted red plottings, respectively). b. TD equilibrium structure. c. TD Raman spectrum. UHF AM1 calculations.

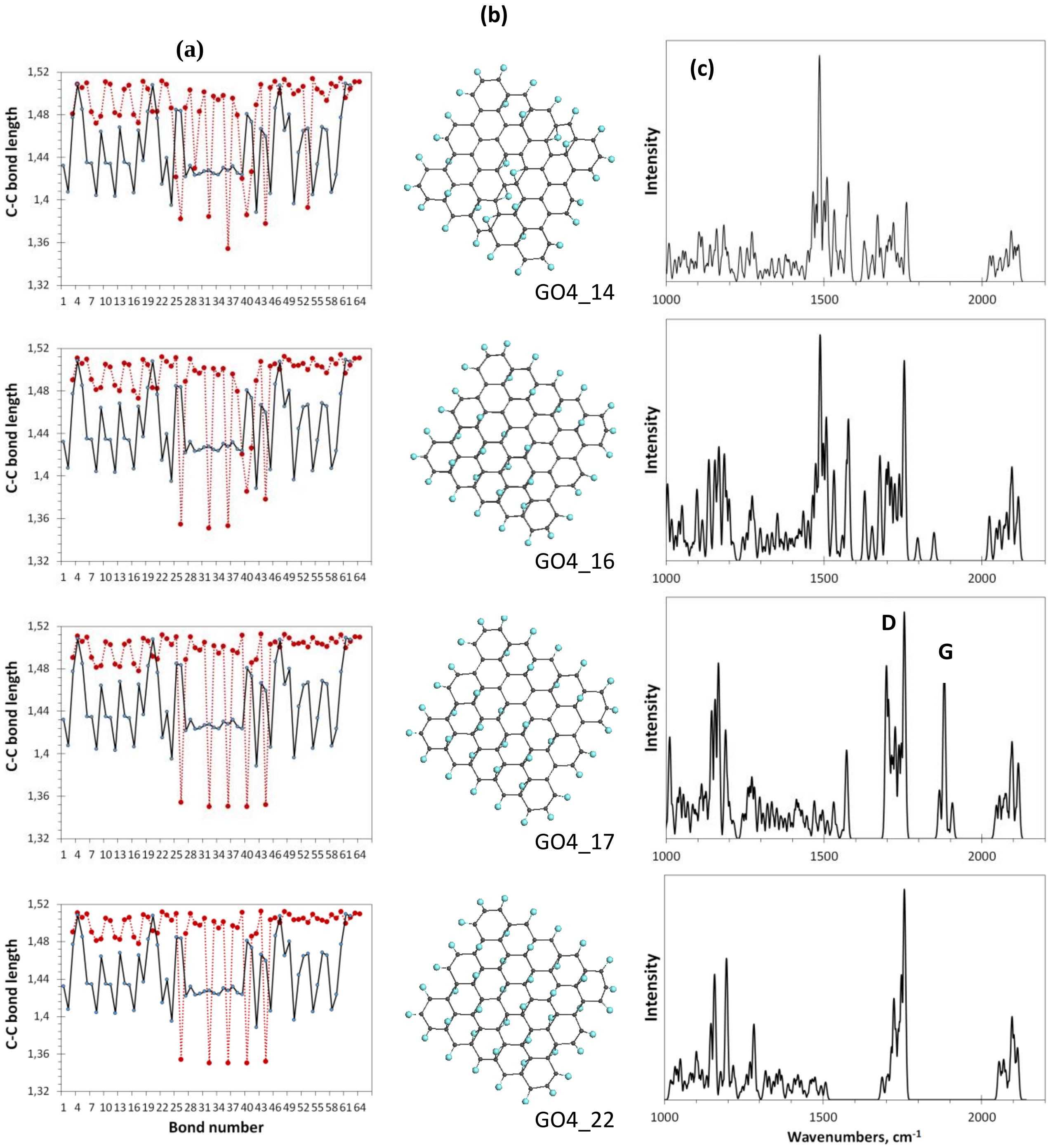

**Figure 8**. Monitoring of the *sp*$^2$-to-*sp*$^3$ C-C bond structure transformation at final steps of the oxidation of basal-plane carbon atoms of GO4. a. TD C-C bond length (reference and current presented with black and dotted red plottings, respectively). b. TD equilibrium structure. c. TD Raman spectrum. UHF AM1 calculations.

Figure 8 present the completion the reaction monitoring. The last steps are followed with severe decreasing of the ACS value, which are zeroing at the 17$^{th}$ step. Simultaneously, the bond length dispersion $\Delta l_{C-C}$ drastically decreases, which causes a remarkable structure ordering of the bonds pool of GO4_17 with respect to preceding ones. The bond distribution reveal two types of bonds, first of which concerns the main massive of single *sp*$^3$C-C ones of 1.50±0.01 Å in length and the second is related to five highly shortened double *sp*$^2$C-C bonds of 1.35±0.001 Å. Such a quasi-threshold ordering of the *sp*$^3$ carbon structure, which accompanies the zeroing of the ACS index, was observed by us in all cases of virtual syntheses of the *sp*$^2$ nanocarbons polyderivatives, such as hydrogenation [41] and fluorination [42] of $C_{60}$ fullerene, hydrogenation [35] and oxidation [36] of graphene domains and so forth [42, 45]. It can be assumed that this effect is because of the accumulated stress of the bond structure does not allow one to bring the addition reaction to 100% in any of the above cases. The termination of the reaction means the vanishing of the ACS, which causes the rearrangement of bonds along the lengths, since the ACS values sharply depend on the latter [43].

As seen in Figure 8c, a dramatic change in the Raman spectra occurs at the 17th step of the addition of the epoxy group to the basal plane simultaneously with the structural transitions described above. The spectrum becomes much more structured and the manifestation of bands **II** and **I** is clearly observed. Further oxidation can be performed manually only by random placing epoxy groups over five remaining double *sp*$^2$C-C bonds. As seen from the figure, the distribution of bonds over lengths of the last-step GO4 (GO4-22), involving a full set of attached epoxy groups, practically does not change in this case. However, the former shortened double *sp*$^2$C-C bonds become single *sp*$^3$ ones and the band **II** disappears. Thus, simultaneous monitoring of the per-step synthesis of the oxide by the C–C bonds pool and Raman spectra allows us concluding that the **I**-**II** doublet in the oxide spectrum is due to the establishment of a balanced structure of highly ordered *sp*$^3$ and *sp*$^2$ C-C bonds, which keeps the carbon skeleton the most stable and undestroyed. The appearance of double C-C bonds, the length of which is shorter than $R_{crit}$ =1,395 Å, which terminates the reaction [43] and provides the undestroyed carbon structure, is the required price. Meeting both requirements is accompanied by an obvious structural ordering of the skeleton along the lengths of the bonds, which manifests itself in a small and almost vanishing dispersion of *sp*$^3$ and *sp*$^2$ C-C bonds, respectively.

Confirmation of what has been said above follows from Figure 9, which shows the vibration spectra of a set of DTs discussed in this paper. The spectrum in panel (a) belongs to the bare domain (5,5)NGr and represents the full spectrum of vibrations provided with a total pool of *sp*$^2$C-C bonds. As seen the spectrum is clearly divided into two parts related to torsions, bendings, breathings, etc (0-1000 cm$^{-1}$) and $\nu$*sp*$^2$**C**-C stretchings (1000-1800 cm$^{-1}$). The spectrum in panel (b) belongs to the same domain in a necklace of 22 oxygen atoms, which corresponds to one of the configurations typical to rGO [32]. As can be seen in the figure, the spectrum preserves two-part view, but two peculiarities appear in the previous stretching part: a group of $\nu$*sp*$^2$**C=O** stretchings at 2100 cm$^{-1}$ and a particular high frequency $\nu$*sp*$^2$**C**-C ones at ~1750 cm$^{-1}$. There is reason to believe that the appearance of this group of $\nu$*sp*$^2$**C**-C stretchings is due precisely to the presence of oxygen atoms in the domain circumference. The appearance of a well-defined gap at ~1700 cm$^{-1}$ should be noted as well, which is due to the *sp*$^2$-to-*sp*$^3$ transition of part of the C-C bonds, which causes their elongation and expected softening.

The spectrum in panel (c) corresponds to the case when the oxidation of the domain (5,5)NGr is terminated because of ACS zeroing. The spectrum is fully reconstructed, particularly in the stretching part over 1000 cm$^{-1}$. It is natural to connect the change with largely expanded stretching pool, which now involves $\nu$*sp*$^3$**C**-**OH,** $\nu$*sp*$^3$**C-O-C,** $\nu$*sp*$^3$**C-C,** $\nu$*sp*$^2$**C-C,** and $\nu$*sp*$^3$**C =O** modes. In accordance with Table 1 we believe that the first two modes fill the band **A**, the pool of $\nu$*sp*$^3$**C-C** ones is responsible for bands **B** and **I,** the latter are highly ordered and low dispersed, the retaining set of shortened $\nu$*sp*$^2$**C-C** ones is responsible for band **II**, and the last $\nu$*sp*$^3$**C =O** modes provide band **III**. A particular attention should be given to the elongation of all the $\nu$*sp*$^3$**C =C** modes of bands **A** and **I** caused by the presence of oxygen atoms with respect to a standard massive of these modes in nanographane (see band **C** in spectrum (d) in the figure [62]), graphane crystal [63], and diamond [66].

The assignment of bands **I** and **II** in the Raman spectrum of GO4_17 to the corresponding features of their vibrational spectra makes us to return to Figure 4b and check, whether the virtually synthesized oxides GO1 and GO3 meet this requirement. Figure 10 shows distributions of the C-C bonds over lengths related to these DTs. Both DTs correspond to incomplete oxidation, which is terminated on the 18 and 38 steps of the reagent placing on the basal plane of the pristine domain, respectively. Due to ACS zeroing [36]. As seen in the figure, in both cases this reaction step is accompanied by the formation of a few shortened $sp^2$C–C bonds of 1.350±0.003 Å in length. It is these low dispersed bonds, which cause the appearance of the band **II** in the virtual spectra of GO1 and GO3, once band **G** in experimental Raman spectra of the GO. As for the arrays of $sp^3$ C-C bonds, they are naturally well structured in both cases and constitute 1,505±0.032 Å and 1,548±0.043 Å, respectively. The same difference in the length and dispersion is observed in the case of GO4 and GO6. Consequently, the gap between bands **C** and **I** of the vibration spectrum of GO3 and GO6 (see Figure 9c) is not formed resulting in the absence of a configured band **I** in the virtual spectra of GO3 (Figure 4b) and GO6 (Figure 6b).

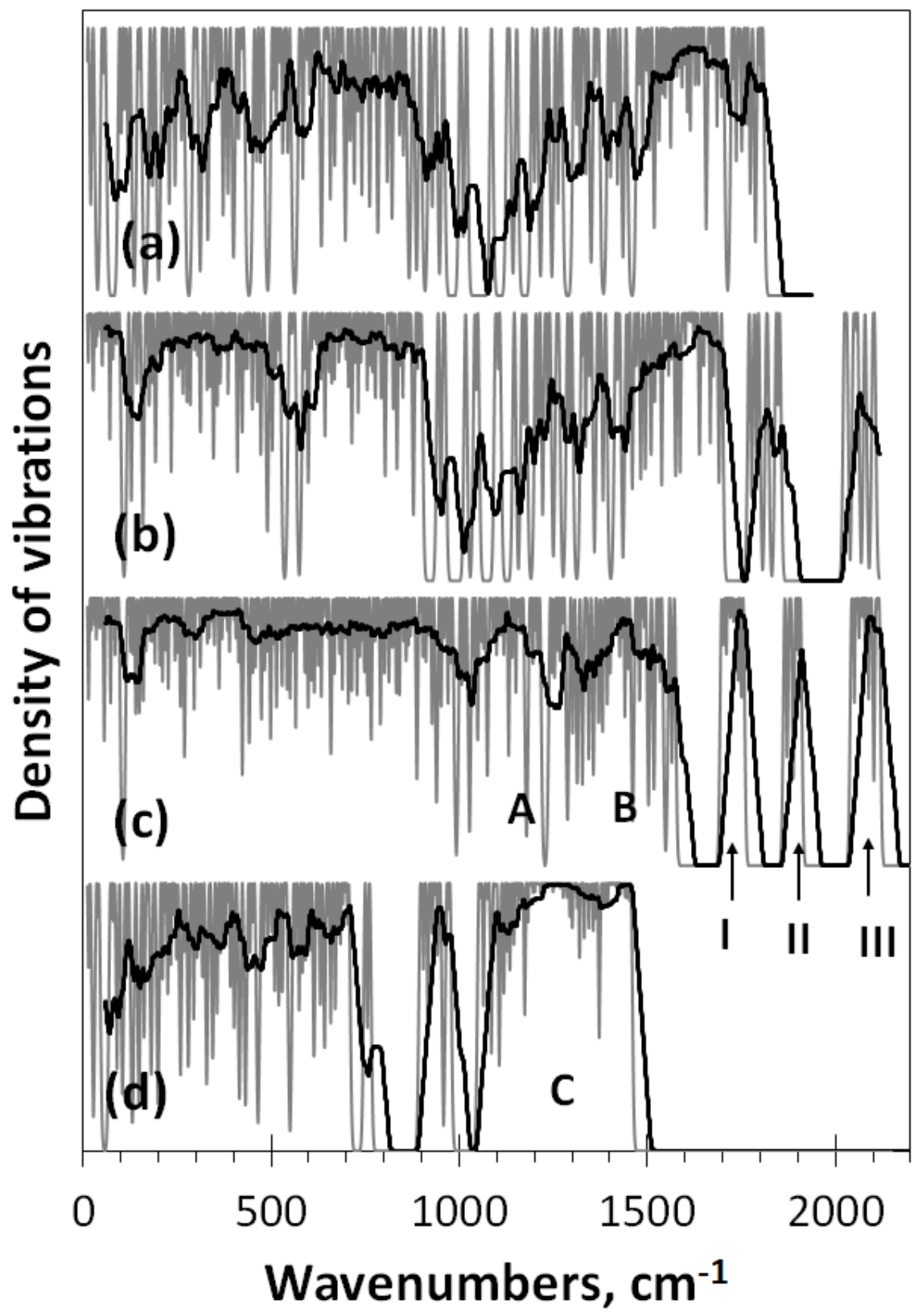

**Figure 9.c** Vibration spectra of bare (5,5)NGr domain $C_{66}$ (a) and its polyderivatives: reduced graphene oxide $C_{66}O_{22}$ (b), graphene oxide GO4_17 $C_{66}O_{39}$ (c), and graphene hydride (nanographane [35]) $C_{66}H_{88}$ (d). UHF (a, b) and RHF (c, d) AM1 calculations.

## 6. Discussion on Digit Twins concept and conclusive remarks

This paper presents the first results of solving a confused problem related to the vibrational spectroscopy of complex molecules from the standpoint of digital twins (DTs) concept. This problem concerns a unique spectral phenomenon - the identity of the Raman spectra of parental and reduced graphene oxides, an explanation for which has not been found to date. In order to solve this problem, it was necessary to obtain answers to a number of questions, among which there are following:

- What is common and what is the difference between the atomic structures of rGO and GO?

- How GO is formed and what secrets does this process hide?
- What general regularities do govern the spectra of both vibrations themselves and their optical counterparts of rGO and GO?
- Are the Raman spectra of rGO and parental GO really identical?
- What information do carry Raman spectra of rGO and GO and how to read them?

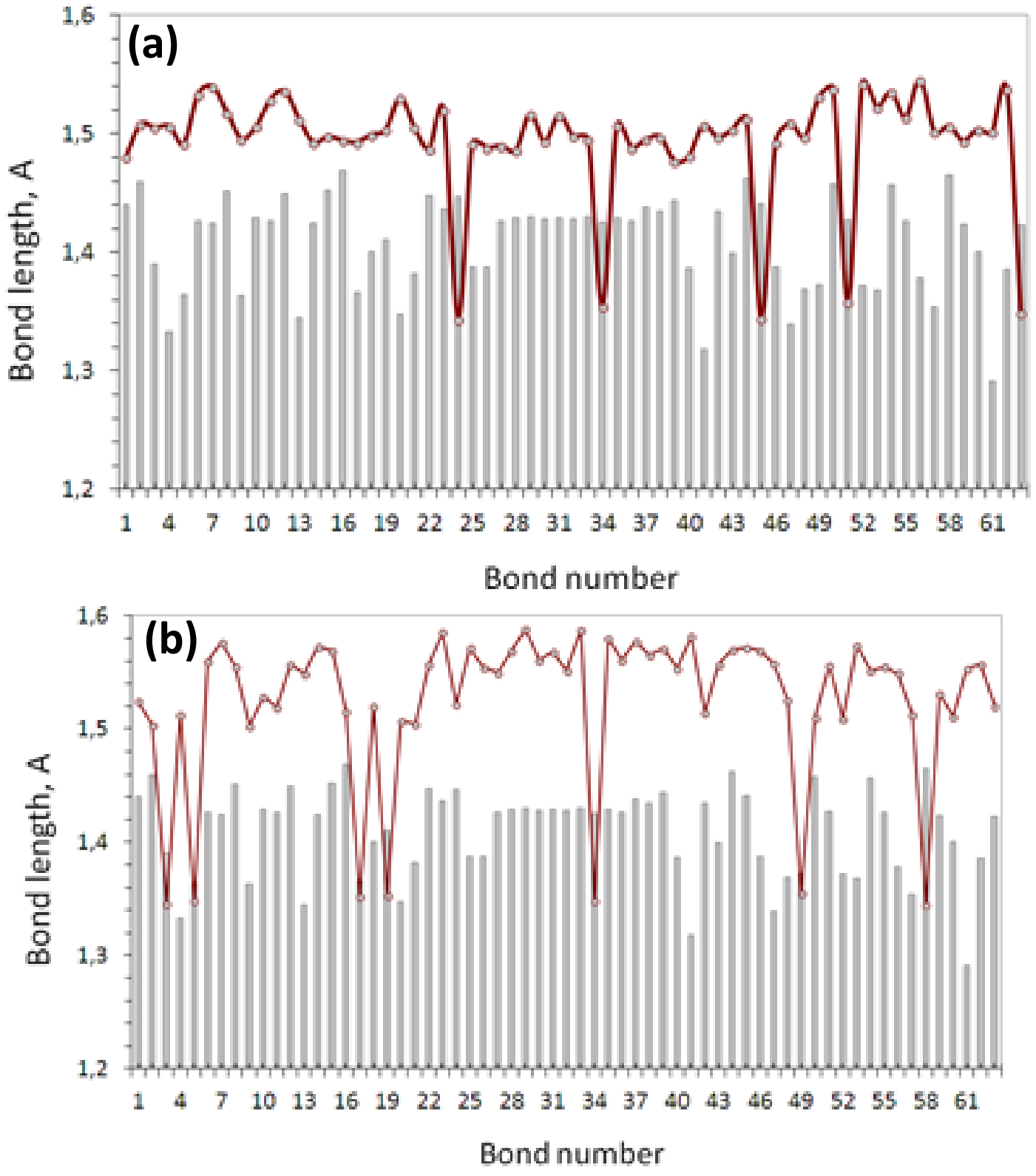

**Figure 10.** The $sp^2$-to-$sp^3$ transformation of the (5, 5) NGr molecule skeleton structure in the course of the successive hetero-oxidant oxidation. The second-stage oxidation related to GO1 (a) and GO3 (b). Light gray histogram is related to the pristine domain (5, 5)NGr.

Naturally, answering these questions requires solving several complex problems, which, in turn, requires a broad program. A wide range of areas of this program, which does not allow its development at the same conceptual level, is an extremely complicating factor, as a result of which no program of this level has been proposed and implemented until now. And only the concept of virtual vibrational spectrometry (VVS) recently proposed by the author [52], which is based and spontaneously merged into the more general concept of DTs one, seems to develop such a program and get answers to all the questions listed above.

In this paper, the DTs concept of virtual vibrational spectrometry is implemented according to the following scheme

$$\textit{Digital twins} \rightarrow \textit{Virtual device} \rightarrow \textit{IT product}.$$

Here, DTs, molecular models by other words, are objects of the study, virtual device is a carrier of a selected software, IT product covers a large set of computational results related to the DTs under different actions in the light of the soft explored. The main feature of the approach is the possibility of practically unlimited object manipulation, restricted only by the possibilities of their perception by a virtual device. A distinctive advantage of the concept is the possibility of a comparative analysis of a huge pool of new information obtained under the same conditions. As it turned out in our case, and which is typical for the concept application to other areas of technology and knowledge (see [67-69] but a few), such a broad coverage not only explains the properties or behavior of real objects, but also reveals hidden, often completely unexpected secrets, which allows you looking at real objects from a fully different side. This has happened in our case with graphene oxides.

As already noted, DTs are independent objects of research, however, their design is not arbitrary, but is determined by the objects of a real problem to be solved. In our case, these objects are rGO and GO. Naturally, DTs must contain all the knowledge gained on a real object, expressed both in the main structural combinations and in their possible modifications. In total, within the framework of the oxide program, which began two years ago, it turned out to be necessary to use 50 DTs. Common to all them was the presence of a basic structural unit (BSU) in the form of a partial or complete polyderivative of a rectangular (n,n) graphene domain, the size of which was determined by the number n of benzenoid units located along its armchair and zizzag edges. The number n was 5, 7, 9, and 11. More than 90% of the DTs structure was based on the (5, 5) domain – (5,5)NGr. Additional bonus of dealing with so big bank of data concerns its full selfconsistency, which allows ignoring unavoidable blue shift of calculated data with respect to experimental ones [70]. Highly efficient HF Spectrodyn provides ~300 $cm^{-1}$ shift for the vibration region of 1000-3500 $cm^{-1}$.

In the chain consisting of graphene domain, rGO, and GO, GO is the most complex and the transition to its consideration obviously requires a thorough analysis of the first two chain members. Accordingly, the first part of the research on the oxide programm was mainly performed for graphene domains [33], while the second was particularly devoted to rGO [32, 34]. The virtual device that made it possible to run this program turned out to be the virtual vibrational spectrometer HF Spectrodyn, which provides the researcher with virtual spectra of IR absorption and Raman scattering of light. The results obtained made it possible to obtain answers to some of the questions formulated above. Thus, one of the important elements of answering the first question is the indisputable establishment of the fact that, rGO presents a wide class of necklaced covalently bonded graphene molecules, each of which is a basic bare graphene domain in a necklace of heteroatoms chemically bonded to it. Answering question 3, it was found that the C-A covalent bonds of various rGOs between edges carbon atoms of the domain and attached heteroatoms A, on the one hand, and the $sp^2$C-C bonds between carbon atoms in the rGOs basal plane, are differently active in optical vibrational spectra. Thus, the C-A bonds are mainly responsible for the IR absorption spectrum, providing their strong variability depending on the rGO origin, while the domain $sp^2$C-C bonds determine the main pattern of the rGO Raman spectra resulting in their empirical standard D-G doublet view, which is practically independent on the rGO origin. It was also found that this typeAs of the rGO Raman spectrum is not a characteristic of an individual molecule but is a signature of the molecules multilayered structure. Moreover, a particular doublet shape strongly depends on linear size of the rGO molecules and is straightway trasformed from a broad-band one to narrow-band with a drastic domination of G band when the size of graphene domain exceeds the free path of graphene optical phonons $L_{ph}$~15 nm. Concerning the hidden secrets of the rGO dynamics, it was possible to find out that the characteristic D-doublet structure of its Raman spectrum is a manifestation of the formation of special dynamic single $sp^3$C-C bonds between atoms located in adjacent layers. Thus, this doublet spectral signature is evidence of the stacked nature of the rGO solid structure.

Coming back to that part of question 3 that is connected with GO, we first have to get an answer to first two questions. The DTs involved in the study discussed in the current paper allows getting a clear answer: GO, like a rGO, is a product of the graphene domain polyderivatization.

However, in contrast to rGO, whose chemical modification is concentrated only in the domain circumference, the GO formation applies to the entire domain, including the carbon atoms of the basal plane. Thus, if in the case of rGO *sp²* hybridization of the electronic structure of carbon atoms is retained not only in the basal plane, but sometimes takes the place for the valence saturated edge atoms as well, the hybridization of the GO atoms becomes *sp³* one throughout the molecule structure. This leads to the replacement of benzenoid planar structures of rGO by cyclohexanoid bent structures of GO, which increases the thickness of the molecule from 0.335 nm to 0.479 nm and this alone leads to an increase in the interlayer distance in solid GO, in contrast to rGO. Thus, the complete oxidation of the domain is accompanied with *sp²*-to-*sp³* transformation of the carbon carcass of the pristine graphene domain.

The formation of an almost planar structure of rGO molecules does not raise any special questions, since the diversity of the rGO chemical compositions, affecting only a few atomic percent of its structure, is in reality easily predictable and is explained by the difference in chemical reactions leading to the formation of a necklace of atoms in each specific case. In the case of GO, the prediction of its structure is more problematic, since there is always the possibility of participation in the process of oxidation of different oxygen-containing groups, on the one hand, and the location of reagents in different parts and on different sides of the basal plane of the original graphene domain. Even 10 years ago, considering this problem [36], we abandoned the simple drawing of a possible hypothetical structure and, realizing the inevitable transformation of the initial *sp²* covalent bonds into *sp³* at each oxidation act, turned to the virtual synthesis of GO using the spin-density algorithm that tracks the chemical activity of the initial carbon atoms by quantitative way. The structures synthesized at that time were used in the present work as one of possible DT sets, which gave an unexpected result: the Raman spectra of two of them (GO1 and GO2) have a characteristic structure of a doublet of **I** and **II** bands located in the same spectral region as in the spectra of rGO. Thus, an affirmative answer to question 4 was obtained, but it remained to find out the reason for the appearance of this pronounced doublet and its replacing with a singlet in a number of cases. It was obvious that the reason lies in the structure of the GO, i.e. in the peculiar features of *sp²*-to-*sp³* transformation of the carbon carcass under the action of various oxidizing agents.

By changing the initial conditions and slightly changing the format of the governing algorithm application, we managed to virtually synthesize 23 GO DTs of various oxidation states thus giving an extended answer to question 2. The results obtained on the basis of these DTs allowed completing the search and get a clear answer to question 3. As in the case of rGO, the GO structure retains two parts associated with the difference in the oxidation processes occurring at the edge atoms of the pristine graphene domain and in its basal plane. In contrast to rGO, all carbon atoms of the GO domain have *sp³* hybridization, in connection with which the atoms on the edges and in the basal plane, it would seem, become similar. However, the difference is preserved, as a result of which the covalent bonds that link heteroatoms to the edge carbons determine the nature and shape of the IR absorption spectrum mainly, while the carbon atoms of the basal plane provide the shape of the Raman spectrum. At the same time, it was shown that the appearance of a doublet Raman spectrum of an GO occurs only when its basal plane is covered with epoxy groups.

Thus, answering the fourth question in the affirmative way, it was shown that the double-band structure of the Raman spectra can indeed be observed in the spectra of both rGO and GO, however, under different conditions. Those related to rGO, once a mystery until recently, was voiced above and addressed to the stacked multilayer structure of the solid. The doulet discloses a pair provided with *sp³*C-C and *sp²*C-C stretchings, the former of wich presents specific dynamical *sp³*C-C bonds formed between carbon atons of adjacent layers. As for GO, a similar pair of *sp³*C-C and *sp²*C-C stretching bands is the result of a *sp²*-to- *sp³* transformation of the carbon carcass, which was not fully completed caused by the oxidation termination. Thus, another secret was discovered, which is inherent in carbon and its unique ability of the $sp^2$-to- $sp^3$ transformation and which consists in the fact that this transformation cannot spontaneously be one hundred percent. This is due to the fact that the rearrangement of the structure, due to the replacement of benzenoid units by cyclohexanoid ones, once accompanied with the inevitable elongation of the initial $sp^3$C-C bonds, leads to a significant mechanical stress of the carbon carcass, maintaining stable continuity of which mandatory requires shortening the lengths of some pristine $sp^3$C-C bonds. However, this feature is accompanied with a significant decrease in the spin density on the corresponding atoms up to complete zeroing, which leads to the termination of the oxidation reaction. The proportion of such shortened bonds is relatively small and amounts to 7–9% of the entire array of bonds, but this amount is completely sufficient to explain the presence of 20–10% unoxidized carbons in the sample array, which has been repeatedly established in practice (see reviews [1-9]), on the one hand, and the appearance of the band **II** related to toughened $sp^2$C-C stretchings.

Having thus obtained answers to all four first questions, we can answer the last one as follows. The characteristic $sp^3$- $sp^2$ D-G doublet structure of the Raman spectrum of rGO tells us that the substance under study consists of necklaced graphene molecules of linear size less than 15 nm that are packed into stacks of different thicknesses. The intensity ratio of the D and G bands should depend on the number of layers in the stack and changes in favor of the D band as this parameter increases (experimental confirmation of the conclusion has bee recently obtained [71]. In turn, the $sp^3$- $sp^2$ D-G doublet structure of the GO Raman spectrum evidences that the $sp^2$-to-$sp^3$ transformation of the carbon carcass has taken place, although not completed to the end due to the oxidation termination, which is manifested by the presence of the G band in this spectrum. Spontaneous cessation of oxidative reactions turns out to be a consequence of the gradual deradicalization of the initial $sp^2$ carbon atoms located in the basal plane.

The proposed analysis of the vibrational spectra of rGO and GO is based on the important concept of the radical nature of bare graphene domains and rGO. Many spears have been broken in discussions that cast doubt on the radical nature of graphene materials. The author would like to hope that the above described study will allow rejecting the last remaining doubts and will force researchers to work with materials familiar to them in a new way.

At the end of writing this article, another feature associated with the chemical element carbon attracted the attention of the author, which, apparently, is worth paying attention to. Surprisingly, carbon and silicon elements behave in nature quite differently. Both elements are adjacent in the column of the periodic table, do not exist in the liquid phase, exhibit a number of similar properties in the gaseous and solid state, but at the same time they have a number of significant differences. Limiting ourselves to a solid body, we note such important points as the absence of a graphite-like structure in silicon, in contrast to the wide distribution of graphite, the relatively easy accessibility of the cubic form of solid silicon, in contrast to the extremely hard one in the case of diamond, and the widespread occurrence of silica in nature, in contrast to the complete absence of carbon oxide. These and other features of the silicon-carbon pair have been the subject of discussion since ancient times. However, the explanation of possible causes began to form relatively recently, which coincided with the emergence of graphenics, which was the stimulus for a wide discussion of this problem. It all started with the phantom of silicene and ended with the understanding that the reason for the discussed differences is the prohibition of the implementation of $sp^2$ hybridization for silicon atoms, in contrast to carbon one [45]. Silicon is the $sp^3$ element only, while solid carbon can exist in all three forms of hybridization: $sp^1$(see about solid acetylene [72]), $sp^2$, and $sp^3$. It was found that a significant increase in the size of silicon atoms and, accordingly, the Si-Si interatomic covalent distances prevents the realization of $sp^1$ and $sp^2$configurations of atoms due to the exceptionally high radicalization of the corresponding atomic compositions, leaving the possibility of stable existence only for the *sp*$^3$ forms.

The basic structural unit of silica is the silicon-oxygen tetrahedron $SiO_4$, whose structure ideally reproduces the *sp*$^3$ hybridization of silicon atoms. In contrast, the carbon-oxygen tetrahedron $CO_4$ is extremely unstable, and even in those cases where its formation is implied, the carbon atom is in *sp*$^2$ hybridization [73]. This circumstance, as well as the obviously very severe conditions for the formation of diamond (40 GP and 960-2000$^0$C) [74], are natural and rule out the possibility of the formation of carbon "silica" or diamond oxide. Unlike *sp*$^3$ diamond, *sp*$^2$graphite is widely distributed; however, graphite oxide does not exist in nature. The reason for this has recently been considered in relation to another natural carbon material – shungite [75]. It was suggested that the formation of an extended *sp*$^2$ material determines the rate of growth of primary carbon lamellae. The graphene lamella grows until its edge carbon atoms are completely terminated by surrounding heteroatoms, the most active of which are OXYGEN ones. The freezing of growth and the continued presence of such a lamella in an oxygen-containing atmosphere is naturally accompanied by the oxidation of atoms lying in the basal plane, accompanied by the *sp*$^2$-to-*sp*$^3$ transformation. However, the presence of water as a reducing agent [76], which usually accompanies graphite deposits, leads to the restoration of the previous *sp*$^2$ configuration caused by the removal of oxygen atoms from the basal plane. During a long time of graphite formation, the processes of oxidation and reduction settle down and the latter wins as more thermodynamically favorable. Graphite oxide, which is absent in nature, is obtained only synthetically by hard oxidation of crushed graphite during a regulated technological cycle [1-9].

Such a long conclusion was needed to draw the reader's attention, firstly, to how important the permissibility of the *sp*$^2$-to-*sp*$^3$ transformation is in the life of solid carbon, and, secondly, to the fact that graphene oxide and reduced graphene oxide are a close pair tightly connected by this transformation to each other. In this work, we will show that it is this unique property of the structural rearrangement of carbon atoms that allows answering the question of the reasons for the identity of the Raman spectra of the parental and reduced graphene oxides.

**Acknowledgements**

The author is thankful to D. Kornilov for supplying with the spectra of the AkKo Lab' GO sample and E.G. Atovmyan for fruitful discussions. This paper has been supported by the RUDN University Strategic Academic Leadership Program.

**Table 1**. General frequencies kits to suitably interpret vibrational spectra of rGOs and GO[1], $cm^{-1}$

| | Spectral areas | | | | | | | | |
|---|---|---|---|---|---|---|---|---|---|
| Experiment[2] | 300-1000 | 1000-1200 | 1200-1300 | 1300-1500 | 1500-1600 | 1600-1700 | 1800-1900 | 2600-3000 | 3000-3600 |
| rGO | *δ op*[4], *δ ip*[5], ρ torsions $sp^2$**C-O-C** and $sp^2$**C-O**H *δ op* $sp^2$**C-C-C**[6] *δ ip*, puckering, ring breathing, *δ* trigonal $sp^2$**C-C-C**[6], collective vibrations of domain atoms[7] | *ν* $sp^2$**C-O-C** in cyclic ether, aggregated cyclic ether and acid anhydride, | *ν* $sp^2$**C-O**H, in lactone, hydroxyl pyran and acid anhydride | *δ ip* $sp^2$**C-O**H, *ν* $sp^2$**C-O-C** in cyclic ether and acid anhydride *δ ip* **O-C**=O in acid anhydride | *δ ip* $sp^2$**C-O**H, *ν* $sp^2$**C-C** | *ν* $sp^2$**C =O** in acid anhydride and lactone, aggregated cyclic ether with lactone pair, pairs of lactones | *ν* $sp^2$**C =O** in *o*-quinone, COOH | *ν* $sp^3$C-**O-H** in COOH *ν* $sp^3$**C-H** | *ν* $sp^3$C-**O-H** |
| GO | *δ op*[4], ρ torsions of $sp^3$**C-O-C** and $sp^3$**C-O**H, *δ op* $sp^3$**C-C-C**[8] *δ ip*, puckering, cyclohexane ring breathing, $sp^3$**C-C-C**[8] | *δ ip* $sp^3$**C-O**H, *ν* $sp^3$**C-OH,** *ν* $sp^3$**C-O-C,** *ν* $sp^3$**C-C**[9] | *δ ip* $sp^3$**C-O-C**, *ν* $sp^3$**C-O-C** In pairs of cyclic ether and lactones. *ν* $sp^3$**C-C**[9] | *δ ip* $sp^3$**C-O**H, *ν* $sp^3$**C-OH,** *ν* $sp^3$**C-O-C,** *ν* $sp^3$**C-C**[9] | *ν* $sp^2$**C-C** | - | *ν* $sp^3$**C =O** | - | *ν* $sp^3$C-**O-H** |
| Virtual data[3] | 300-990 | 1280 | 1410-1550 | 1560-1660 | 1800-1900 | - | 1950- 2100 | - | 3420 |

[1] Greek symbols δ and ν mark the molecule bendings and stretchings, respectively
[2] The assignment of frequencies in the experimental spectra is based on the papers [54-57].
[3] Obtained in the current study
[4] Out-of-plane bendings
[5] In plane bendings
[6] Benzene molecule data [58]
[7] Virtual data for nanographene [59]
[8] The author approximated suggestions for GO
[9] $sp^3$**C-C** stretchings cover regions of 1